\documentclass[10pt]{article}

\usepackage{amsmath}
\usepackage{amssymb}
\usepackage{textcomp}
\usepackage{mathrsfs}
\usepackage{graphicx}

\usepackage{authblk} 
\usepackage{cite} 

\usepackage[dvips]{epsfig}
\usepackage{psfrag}

\usepackage[margin = 2.5 cm]{geometry}

\begin{document}

\title{Analyzing effective models: An example from JAK/STAT5 signaling}

\date{}

\author[1]{Martin Peifer}
\author[2,3]{Jens Timmer}
\author[4]{Christian Fleck}

\affil[1]{Department of Translational Genomics, University of Cologne, Germany}
\affil[2]{Institute for Physics, University of Freiburg, Germany}
\affil[3]{BIOSS Centre for Biological Signalling Studies, University of Freiburg, Germany}
\affil[4]{Laboratory for Systems and Synthetic Biology, Wageningen University, 6703 HB Wageningen, The Netherlands}

\maketitle

\section{Abstract}
In systems biology effective models are widely used due to the complexity of biological system. They result from a coarse-graining process which employs specific assumptions. Frequently one does not start with a model taking all details into account and then performs a coarse-graining process, but rather one starts right away with the effective equations and often the underlying assumptions remain hidden or unclear. We exemplify the analysis of an effective model by analyzing a time delay equation for the JAK/STAT5 signaling pathway and show how one can avoid wrong conclusions and obtain a deeper understanding of the biological system .
By analyzing the assumptions leading to a coarse-grained model one might be able to gain new insight into the involved biological processes. Further, the compliance of the model with experimental data can be considered as a validation of the assumptions made in the derivation of the mathematical equations.      


\section{Introduction}
Biological processes such as, e.g., signal transduction pathways, often involve many different time-scales ranging from a $\mu$s to hours and also span often several length scales. A prominent example for an event on a short time and length scale involved in signal transduction is the binding of a ligand to a receptor or protein-kinase interaction and the subsequent phosphorylation \cite{Pawson:2005bx,Brivanlou:2002uk}. Each of these is a complex phenomenon itself, e.g., the binding of a ligand to a receptor is governed by different physical interactions, van der Waals, electrostatic, etc., but also by the mobility of the ligand and the structure of the proximity of the receptor \cite{Grima:2006to}. Clearly, models including all these details would be intractable. Rather, one refers to coarse-grained or simplified effective models reflecting the level of detail necessary to study a specific problem and avoid considering all details of the underlying complex process. Another well known example of coarse-grained models are Langevin equation where the influence of the water molecules on collodial particles is comprised by a stochastic force \cite{Snook:2007ux}.  By using coarse-grained models fundamental design principles are not masked by unnecessary details \cite{Gonze:2011ki}. Alternatively, employing effective models might be necessary if not all components of the system are accessible by experiments. Examples of the effective treatment of biological processes are the synthesis  and the degradation of proteins which involve many sub-steps, such as transcription factor binding, binding of the RNA polymerase, etc., or several ubiquitination steps and translocation to the proteasome, etc \cite{alberts2008molecular}. In all these examples sub-processes are condensed into a single rate which appear in effective models. Another common reduction of sub-processes is their description by specific mathematical functions. Prominent examples are the Michaelis-Menten or Hill functions often used to describe saturating sub-processes by coarse-grained models \cite{Weiss:1997wq}.  Sub-processes can also introduce time delays into the effective mathematical description of a biological system. One of the most famous example of a dynamical system with time delays was suggested by Mackey and Glass \cite{mackey1977,glass1979}, where nonlinear differential-delay equations describe physiological control systems.  Other examples occur in population models \cite{may1975,gurney1980}, infection models \cite{Culshaw:2000p11597,Nelson:2002p11550,Nelson:2000p11549}, signal transduction networks \cite{srividhya2007,swameye2003}, and gene regulation \cite{jensen2003,bratsun2005}. Although effective models are inevitable their use should be a matter of great care, because they result from a coarse-graining process which involves certain assumptions. Of course, in most case one does not start with a model taking all necessary details into account and then performs a coarse-graining process to derive the relevant effective model for the problem at hand.  Rather one starts right away with the effective model or equation and very often the underlying assumptions remain hidden or unclear. In systems biology effective models are widely used as otherwise one would be overwhelmed by the amount of details of a complex biological system. However, employing such models blindly without understanding the range of its validity clearly impairs reliable statements about the system under investigation. Further, models for dynamic biological systems are often tested against quantitative experimental data. If a model reproduces the behavior of a system it is understood that the model sufficiently reflects the dominant dynamics of the system under consideration. This must be then also correct for the implicit assumptions underlying the employed effective model. In turn this means that by analyzing the assumptions leading to the effective model one might be able to gain a deeper understanding about the involved biological processes. In this paper, we exemplify the analysis of an effective model and how one can obtain a deeper understanding of the biological system by analyzing a time delay equation for the JAK/STAT5 signaling pathway previously presented in \cite{swameye2003}.

\section*{Results and Discussion}

\subsection*{The delay differential equation for the JAK/STAT5 signaling pathway as an example for an effective model}
The mathematical model of the JAK/STAT presented by Swameye et al. \cite{swameye2003} presupposes the binding of the ligand erythropoietin (Epo) to the Epo receptor (EpoR) located at the cell membrane. This results in an activation of the receptor (via cross-phosphorylation of the JAK proteins) and a subsequent phosphorylation of the STAT5 molecule. Two phosphorylated STAT5 proteins form a homodimer which enters the cell nucleus, where it stimulates the transcription of target genes. Then the dimers dissociate, are dephosphorylated, and relocated back to the cytoplasm. A delay $\tau$ represents the time the STAT5 proteins reside in the nucleus. This time-delay is introduced to lump all nuclear processes into a single variable. This biochemical reaction scheme can be visualized by
\begin{eqnarray}
\label{reaction}
\mbox{EpoR + STAT5}&\xrightarrow{k_{1}}& \mbox{EpoR + pSTAT5}\nonumber\\
\mbox{pSTAT5+pSTAT5}&\xrightarrow{k_{2}}& \mbox{(pSTAT5)}_2\qquad~\nonumber\\
\mbox{(pSTAT5)}_2&\xrightarrow{k_{3}}& {}_{\mathrm{N}}\mbox{(pSTAT5)}_2\quad~~\nonumber\\
{}_{\mathrm{N}}\mbox{(pSTAT5)}_2&\xrightarrow[\mathrm{Delay}\; \tau]{k_{4}}& \mbox{2 STAT5}\qquad~~~\;,
\end{eqnarray}
where pSTAT5 represents phosphorylated STAT5 in the cytoplasm, ${\rm (pSTAT5)}_2$ the STAT5 dimer, and
the STAT5 dimer inside the nucleus is represented by ${}_{\bf N}{\rm (pSTAT5)}_2$. Reaction rates are represented by $k_1,\ldots,k_4$ and the time-delay by $\tau$. In the study of Swameye et al. \cite{swameye2003} the authors suggested the following model:
\begin{eqnarray}
\label{eq:1}
\dot{x}_1(t)&=&-k_1x_1(t)\mathrm{EpoR}(t)+k_4x_3(t-\tau)\nonumber\\
\dot{x}_2(t)&=&~~k_1x_1(t)\mathrm{EpoR}(t)-k_2x_2(t)^2\nonumber\\
\dot{x}_3(t)&=&~~\frac{k_2}{2}x_2(t)^2-k_3x_3(t)\nonumber\\
\dot{x}_4(t)&=&k_3x_3(t)-k_4x_3(t-\tau) \;.
\end{eqnarray}
Here, the concentration of cytoplasmic unphosphorylated STAT5 is represented by $x_1$,  whereas $x_2$ denotes the phosphorylated STAT5. Moreover, $x_3$ describes the concentration of the dimer and $x_4$ is the nuclear STAT5. The concentration of the activated receptor sites is given by $\mathrm{EpoR}(t)$. Eq.~(\ref{eq:1}) is an example of an effective description of a biological process. Based on this model the authors found from a fit to experimental data that the import rate $k_3$ and the export rate $k_4$ are very similar, in particular, $k_3=0.1066~\mathrm{min}^{-1}(+0.003/-0.022)$ and $k_4=0.10658~\mathrm{min}^{-1}(+0.00016/-0.0024)$. The question arises whether it is valid to draw the obvious conclusion that the import and the export rates for the STAT5 molecules are equal. In order to answer this question we analyze the underlying assumptions of the model. To this end we translate the reaction scheme given by  Eq.~(\ref{reaction}) into a mathematical model, assuming that all reactions inside the nucleus can be summarized by series of reactions. Furthermore, we assume that due to the size of the nucleus, any delay-distribution arising from diffusion can safely be neglected.  In the next section we derive a set of differential equation for a simple two-compartment system.

\subsection*{Delays as an approximation for a series of reactions}
\label{sec:series}
 
Let us assume that a compartment B is divided into several sub-compartments $B_i$ each representing a specific state of the molecules in B. The transition from state $B_i$ into $B_{i+1}$ occurs with rate $\beta_i$, where we assume that the back-reaction rates are significantly smaller than $\beta_i$. This allows us to neglect any reaction from $B_{i+1}$ to $B_i$, such that we clearly assign a direction towards the last sub-compartment $B_n$. The sub-division process of B is graphically illustrated in Fig.~\ref{fig:1}. Note that if the rates $\beta_i$ are very heterogeneous, the behavior of the system is dominated by the smallest rate, as shown in \cite{epstein-revphyschem92}. However, to simplify the discussion we set $\beta_i=\beta$. These considerations lead to the following system of differential equations:
\begin{figure}[t]
\begin{center}
\includegraphics[width=0.5\textwidth]{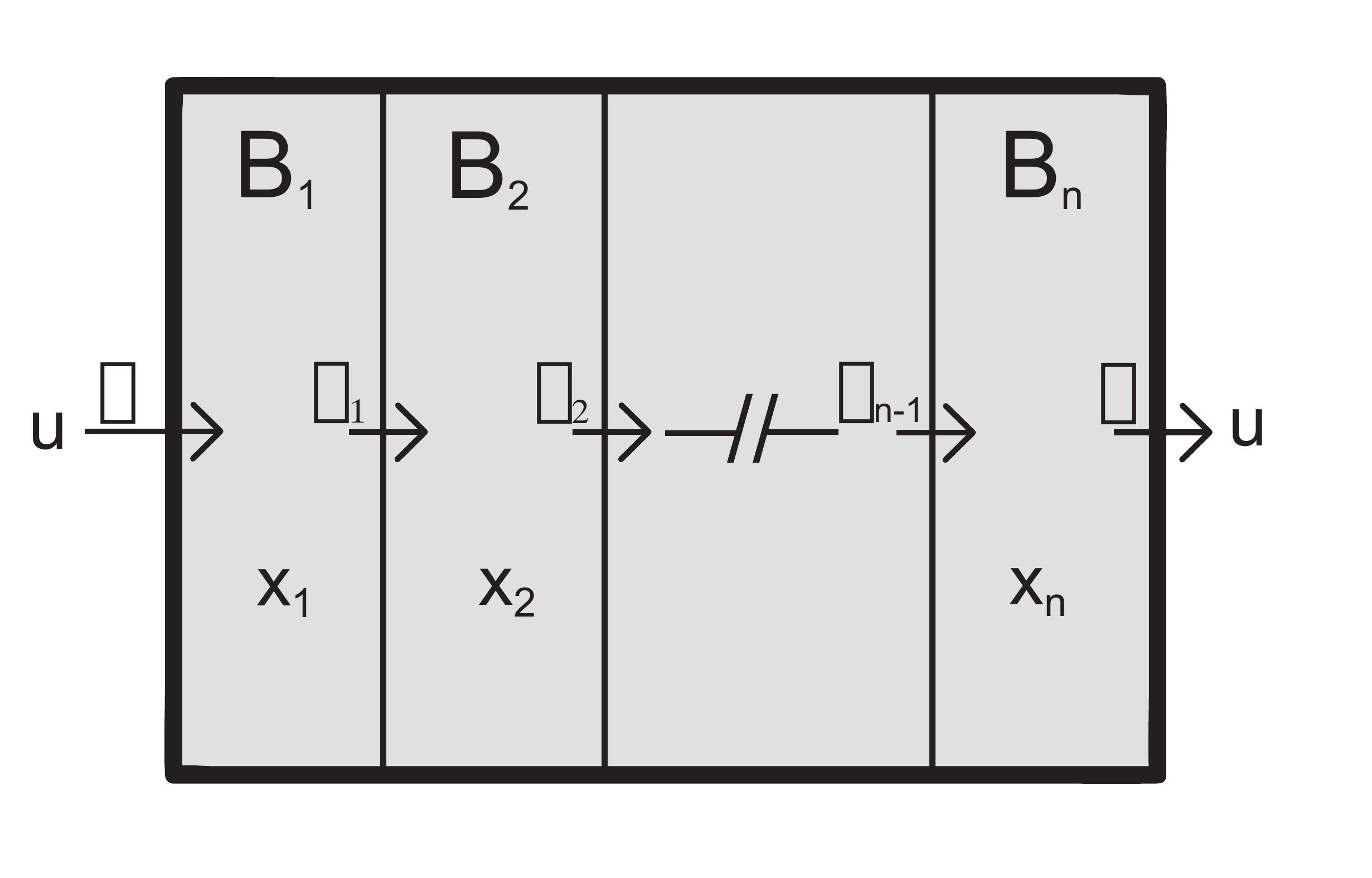} 
\caption{\label{fig:1} Resolving the delay. Compartment B consists of $n$ sub-compartments. Each compartment represents a state of the protein, whereas the transition between compartment $B_i$ and $B_{i+1}$ occurs with rate $\beta_i$ ($\beta_n=\delta$).}
\end{center}
\end{figure}

\begin{eqnarray}
\label{eq:2}
\dot{u}(t)    &=&-\alpha u(t)+\delta x_n(t)\nonumber\\
\dot{x}_1(t)&=&~~\alpha u(t)-\beta x_1(t)\nonumber\\
\dot{x}_2(t)&=&~\beta x_1(t)-\beta x_2(t)\nonumber\\
\label{eq:3}
&\vdots&\nonumber\\
\dot{x}_n(t)&=&\beta x_{n-1}(t)-\delta x_n(t)\;,
\end{eqnarray}
where $x_i$ is the concentration of state $i$ in sub-compartment $B_i$. We further assume that the initial concentration in all sub-compartments is zero. To obtain an equation similar to Eq.~(\ref{eq:1}), we have to relate each $x_i(t)$ to the input $u(t)$ of compartment B. Then, the total
concentration in B is given by the sum over the sub-compartments $x(t)=\sum_{i=1}^n x_i(t)$ which then turns the $n$ differential equations into a single equation containing $x(t)$ and $u(t)$. To this end, we perform a Laplace-transformation of Eq.~(\ref{eq:2}), which leads to
\begin{eqnarray}
\label{eq:3}
s\mathcal{L}(x_1)&=&~\alpha\mathcal{L}(u)-\beta\mathcal{L}(x_1)\nonumber\\
s\mathcal{L}(x_2)&=&\beta\mathcal{L}(x_1)-\beta\mathcal{L}(x_2)\nonumber\\
&\vdots&\nonumber\\
s\mathcal{L}(x_n)&=&\beta\mathcal{L}(x_{n-1})-\delta\mathcal{L}(x_n) \;,\label{eq:4}
\end{eqnarray}
where $\mathcal{L}(\cdot)$ denotes the Laplace-transformation of the corresponding function. Recursively, solving these equations, we arrive at
\begin{eqnarray}
\label{eq:4}
\mathcal{L}(x_n)&=&\frac{\alpha\beta^{n-1}}{(s+\beta)^{n-1}(s+\delta)}\mathcal{L}(u)\;.
\end{eqnarray}
Summing over the sub-compartments and performing the Laplace back transformation finally yields
\begin{eqnarray}
\dot{u}(t)&=&-\alpha u(t)+\alpha\int\limits_0^{t}\,K_n(t-t^{\prime})\,u(t^{\prime})\,dt^{\prime}\nonumber\\
\label{eq:5}
\dot{x}(t)&=&~~\alpha u(t)-\alpha\int\limits_0^{t}\,K_n(t-t^{\prime})\,u(t^{\prime})\,dt^{\prime}.
\end{eqnarray}
Thus the sequence of reactions shown in Fig.~\ref{fig:1} results in a system of integro-differential equations, where the integration kernel is given by 
\begin{eqnarray}
\label{eq:6}
K_{n}(t)&=&\frac{\delta e^{-\delta t}}{(1-\tau\delta/(n-1))^{n-1}}\left[1-\frac{\Gamma(n-1,(1-\tau\delta/(n-1))(n-1)t/\tau)}{(n-2)!}\right],
\end{eqnarray}
where we set $\beta=(n-1)/\tau$ and the function $\Gamma$ is the incomplete gamma function defined by: $\Gamma(a,x)=\int^{\infty}_xt^{a-1}e^{-t}dt$ \cite{gradshteyn}.

\begin{figure}[t]
\begin{center}
\includegraphics[width=0.5\textwidth]{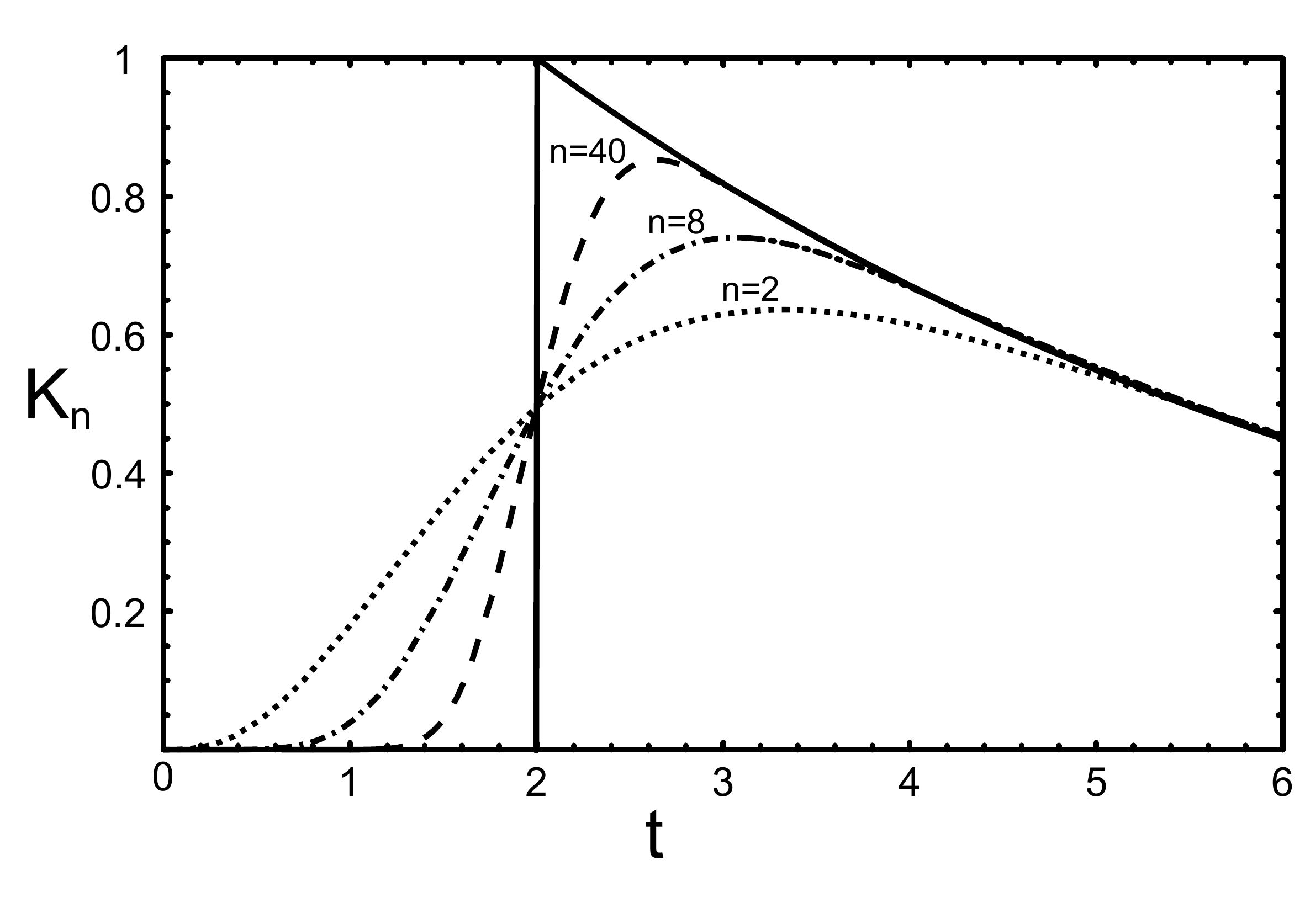} 
\caption{\label{fig:2} The integration kernel $K_n(t)$ for the series of reactions as given by Eq.~(\ref{eq:7})  for different numbers of sub-compartments $n$. In all shown cases we set $\alpha=1$, $\delta=2$ and $\tau=2$ . The solid black line shows $K_{\infty}(t)$, the limiting case $n \to\infty$, given by Eq.~(\ref{eq:7}). }
\end{center}
\end{figure}

Before we employ our results to the JAK-STAT signaling case, we discuss some limiting cases of $K_n(t)$. First, for a large number of sub-compartments $n$ the integration kernel converges to 
\begin{eqnarray}
\label{eq:7}
K_{\infty}(t)&=\lim\limits_{n\to\infty}K_{n}(t)=&\delta\theta(t-\tau)e^{-\delta(t-\tau)}\;.
\end{eqnarray}
\noindent
In Fig.~\ref{fig:2} we present $K_n(t)$ for different number of sub-compartments $n$. It can be seen that already for $n=40$ the limit kernel $K_{\infty}(t)$ provides a good approximation for $K_n(t)$. For smaller $n$ the difference between $K_n(t)$ and $K_{\infty}(t)$ becomes pronounced for $t-\tau<\delta^{-1}$. For $t-\tau\gg\delta^{-1}$ the kernel is dominated by the exponential decay and the difference between $K_n(t)$ and the limit kernel $K_{\infty}(t)$ is negligible small. This means that for small $\tau$ and large $\delta$ even the case of having only a few reactions in B the process can be modeled with sufficient accuracy by using $K_{\infty}(t)$ instead of $K_n(t)$. In the opposite case, where $\tau$ is large or $\delta$ is small it is necessary to use $K_n(t)$. If we further set $\beta=\delta$, the integration kernel given in Eq.~(\ref{eq:7}) simplifies to 
\begin{eqnarray}
\label{eq:8}
K_n(t)&=&\frac{n^ne^{-nt/\tau}t^{n-1}}{\tau^n(n-1)!} \;,
\end{eqnarray}
which is a gamma distribution. The fact that a linear chain such as that in Eq.~(\ref{eq:2}) is equivalent to a delay differential equation with a gamma distributed delay is a known result \cite{macdonald}. We would like to stress that in our model the rate $\beta$ is conceptually different from the rates $\alpha$ and $\delta$.  While $\alpha$ and $\delta$ represent the import and export rate, resp., $\beta$ is related to unknown reactions inside B. The steady state  of Eq.~(\ref{eq:5}) is given by (cf. Appendix \ref{appA}):
\begin{eqnarray}
\label{eq:9}
u^*=\lim_{t\to\infty}\lim_{n\to\infty}  u(t)&=&\frac{u_0 \delta}{\alpha+\delta+\alpha\delta \tau}\;,
\end{eqnarray}
where $u_0$ denotes the initial concentration of $u$ and the limit $n\to\infty$ is motivated by the observation that the dynamics is quite insensitive to $n$. The steady state value of $u$ is decreased compared to a system without delay due to the extra factor $\alpha\delta \tau$ in the denominator which reflects the storage of molecules in the reaction chain.

\subsection*{Analysis of the JAK-STAT system}
\label{sec:JAK}

We are now equipped to analyze the delay system given by Eq.~(\ref{reaction}). Taking the same approach as above we obtain for the JAK-STAT signaling pathway the following integro-differential equations:
\begin{eqnarray}
\label{eq:10}
\dot{x}_1(t)&=&-k_1x_1(t)\mathrm{EpoR}(t)+k_3 k_4\int\limits_0^{t}\theta(t-t'-\tau)e^{-k_4(t-t'-\tau)}x_3(t')~dt'\nonumber\\
\dot{x}_2(t)&=&~~k_1x_1(t)\mathrm{EpoR}(t)-k_2x_2^2(t)\nonumber\\
\dot{x}_3(t)&=&~~\frac{k_2}{2}x_2^2(t)-k_3x_3(t)\nonumber\\
\dot{x}_4(t)&=&~~k_3x_3(t)-k_3 k_4\int\limits_0^{t}\theta(t-t'-\tau)e^{-k_4(t-t'-\tau)}x_3(t')~dt'.
\end{eqnarray}
The resulting integro-differential equation is difficult to handle; to simplify the effective equations further we employ the fact that $x_3(0)=0$ and assume $\sup_{t \in [0,\infty)} |\dot{x}_3(t)|/k_4\ll 1$, which results in (cf. Appendix \ref{appB}):
\begin{eqnarray}
\label{eq:11}
\dot{x}_1(t)&=&-k_1x_1(t)\mathrm{EpoR}(t)+k_3 \theta(t-\tau)x_3(t-\tau)\nonumber\\
\dot{x}_2(t)&=&~~k_1x_1(t)\mathrm{EpoR}(t)-k_2x_2(t)^2\nonumber\\
\dot{x}_3(t)&=&~~\frac{k_2}{2}x_2(t)^2-k_3x_3(t)\nonumber\\
\dot{x}_4(t)&=&~~k_3x_3(t)-k_3 \theta(t-\tau)x_3(t-\tau).
\end{eqnarray}
It is important to recognize that the rate $k_4$ at which STAT5 is exported from the nucleus does not appear explicitly in these equations, a direct consequence of the assumption that the export rate is fast compared to the rate of change of the STAT5 dimer concentration in the cytoplasm. In addition, the dynamic of the nuclear STAT5, $x_4$, is completely determined by the dynamic of the cytoplasmic STAT5 dimers, $x_3$. This might be counter-intuitive and will not be generally true,  but results from the assumptions underlying Eq. (\ref{eq:11}) and is a consequence of incorporating all reactions into a single delay.

We derived starting from a system of equations in which all reactions steps inside the nucleus explicitly appear, a set of effective equations, Eq. (\ref{eq:11}), which are almost equal to the equations used by Swameye et al. \cite{swameye2003}, given by Eq. (\ref{eq:2}). The subtle but important difference is that the export rate $k_4$ does not appear explicitly in Eq. (\ref{eq:11}) as a consequence of our derivation, whereas it is still present in model Eq. (\ref{eq:2}). However, by parameter estimation using quantitative experimental data Swameye et al. found $k_3=k_4$, which can be regarded as an experimental validation of the assumptions underlying Eq.  (\ref{eq:11}). In particular, this means that the processes inside the nucleus can be described as a linear forward reaction process. Moreover, we can conclude that the export process is, compared to the other processes involved, a rather fast process, since we required this in order to achieve Eq.~(\ref{eq:11}) from Eq.~(\ref{eq:10}). Our analysis shows that the seemingly obvious conclusion that the export and the import rate of the STAT5 protein from and into the nucleus, resp., are approximatively equal is not valid and would have been drawn due to a misunderstanding of the equations  Eq.~(\ref{eq:1}). 


\section{Conclusions}

Many biological reactions consist of intermediate steps introducing a complexity which is difficult to handle explicitly. These intermediate steps are often subsumed in effective models such as delay differential equations. We showed that it is necessary to introduce delays with great care, since intuitive modeling might lead to a set of equations which are hard or even not possible to interpret from a mechanistic point of view. We argue that it is important to make the assumptions underlying an effective model explicit in order to arrive to valid statements about the systems behavior. As a consequence, being aware of the requirements leading to the effective model yields extra insight into the system. The compliance of the model with experimental data can be considered as a validation of the assumptions made in the derivation of the mathematical equations. 

As an example of a system of effective equations successfully employed in systems biology we chose the JAK/STAT5 signaling cascade proposed in \cite{swameye2003}. One of the important results for JAK/STAT5 signaling obtained in \cite{swameye2003} is the STAT5 molecule is involved in many subsequent signaling steps inside the nucleus. Instead of guessing the effective equation we derived the delay differential equation through a sequence of simplification steps based on certain assumptions, finally arriving to a similar system of delay differential equations suggested in \cite{swameye2003}. By this procedure we could make the assumptions underlying effective delay differential equations explicit. In particular, since we can describe the action of STAT5 in the nucleus by a linear directed series of reactions it follows that there are no significant feedbacks or back reactions in the signaling cascade inside the nucleus. Moreover, we show that the model does not allow any statements about the export rate $k_4$ of the STAT5 molecules from the nucleus. The considered effective model is well supported by quantitative data and hence this suggests that the conditions for the coarse-graining are fulfilled.


\section*{Acknowledgements}
This work was supported by FP7 CancerSys Project (HEALTH-F4-2008-223188) and BMBF Project FRISYS (0313921).

\begin{appendix}
\section{Steady state of the series of reaction model}
\label{appA}
Starting point for the calculation of the steady state is Eq. (\ref{eq:6}) using the limiting kernel for $n\to\infty$:
\begin{eqnarray}
\label{eq:24}
\dot{u}(t)&=&-\alpha u(t)+\alpha\delta\int\limits_0^{t}\theta(t-\tau-t^{\prime})e^{-\delta(t-\tau-t^{\prime})}\,u(t^{\prime})\,dt^{\prime}.
\end{eqnarray}
Taking the derivative with respect to time results in:
\begin{eqnarray}
\label{eq:25}
\ddot{u}(t)+\alpha\dot{u}(t)&=&\alpha\delta u(t-\tau)-\alpha\delta^2\int\limits_0^{t}\theta(t-\tau-t^{\prime})e^{-\delta(t-\tau-t^{\prime})}\,u(t^{\prime})\,dt^{\prime}.
\end{eqnarray}
Using Eq. (\ref{eq:24}) to replace the integral yields:
\begin{eqnarray}
\label{eq:26}
\ddot{u}(t)+(\alpha+\delta)\dot{u}(t)&=&\alpha\delta[u(t-\tau)-u(t)].
\end{eqnarray}
Next we expand $u(t-\tau)$ in a Taylor series in $\tau$ giving rise to:
\begin{eqnarray}
\label{eq:27}
\ddot{u}(t)+(\alpha+\delta)\dot{u}(t)&=&\alpha\delta\sum_{n=1}^{\infty}\frac{u^n(t)}{n!}(-\tau)^n.
\end{eqnarray}
For $t\le\tau$ Eq. (\ref{eq:24}) can be solved exactly with the result: $u(t)=u_0e^{-\alpha t}$. This can be exploited in the following way.
We integrate Eq. (\ref{eq:27}) over the interval $[\tau,t]$ leading to:
\begin{eqnarray}
\label{eq:28}
\dot{u}(t)+(\alpha+\delta)u(t)=\dot{u}(\tau)+(\alpha+\delta)u(\tau)+\alpha\delta\sum_{n=1}^{\infty}\left(\frac{u^{n-1}(t)-u^{n-1}(\tau)}{n!}(-\tau)^n\right).
\end{eqnarray}
Using the $u(\tau)=u_0e^{-\alpha\tau}$ we obtain for the steady state $u^*$:
\begin{eqnarray}
\label{eq:29}
(\alpha+\delta)u^*=\delta u(\tau)-\alpha\delta\tau u^*+\delta(e^{\alpha\tau}-1)u(\tau).
\end{eqnarray}
which results into Eq. (\ref{eq:9}) for the steady state.

\section{Simplifying the integro-differential equation}
\label{appB}
We use the identity $k_4e^{-k_4(t-t'-\tau)}=d/dt'\,e^{-k_4(t-t'-\tau)}$ to rewrite the integral from Eq. (\ref{eq:10})
\begin{eqnarray*}
k_3 k_4\int\limits_0^{t}\theta(t-t'-\tau)e^{-k_4(t-t'-\tau)}x_3(t')~dt'=k_3\int\limits_0^{t-\tau}x_3(t')\frac{d}{dt'}e^{-k_4(t-t'-\tau)}~dt'.
\end{eqnarray*}
Performing partial integration yields:
\begin{eqnarray*}
k_3\int\limits_0^{t-\tau}x_3(t')\frac{d}{dt'}e^{-k_4(t-t'-\tau)}~dt'=k_3 x_3(t-\tau)-k_3e^{-k_4(t-\tau)}\int\limits_0^{k_4(t-\tau)}\frac{\dot{x}_3(u/k_4)}{k_4}e^{u}~du.
\end{eqnarray*}
If $\sup_{t \in [0,\infty)} |\dot{x}_3(t)|/k_4\ll 1$ holds we can neglect the integral on the right hand side.

\end{appendix}

\bibliographystyle{unsrt}

\begin{small}
	\bibliography{EffectiveModels}
\end{small}

\end{document}